# 5-minute Solar Oscillations and Ion Cyclotron Waves in the Solar Wind

A. Guglielmi[1] • A. Potapov[2] • B. Dovbnya[3]

[1]Institute of Physics of the Earth of the Russian Academy of Sciences, Moscow, Russia
[2]Institute of Solar-Terrestrial Physics of Siberian Branch of the Russian Academy of Sciences, Irkutsk, Russia
[3]Geophysical Observatory Borok of Institute of Physics of the Earth of the Russian Academy of Sciences, Borok, Yaroslavl Region, Russia

**Abstract.** In the present paper we study impact of the photospheric 5-minute oscillations on the ion cyclotron waves in the solar wind. We proceed from the assumption that the ion cyclotron waves in solar wind are experiencing modulation with a characteristic period of 5 minutes under the influence of Alfvén waves driven by photospheric motions. The theory presented in our paper predicts a deep frequency modulation of the ion cyclotron waves. The frequency modulation is expected mainly from variations in orientation of the IMF lines. In turn, the variations in orientation are caused by the Alfvén waves, propagating from the Sun. To test the theoretical predictions we have analyzed records of the ultra-low-frequency (ULF) geoelectromagnetic waves in order to find the permanent quasi-monochromatic oscillations of natural origin in the Pc1–2 frequency band (0.1–5 Hz), the carrier frequency of which varies with time in a wide range. As a result we found the so-called "serpentine emission" (SE), which was observed in Antarctic at the Vostok station near the South Geomagnetic Pole. The permanency, range of frequencies, and the deep frequency modulation of SE correspond to the qualitative properties of ion cyclotron waves in the solar wind. Clearly expressed 5-minute modulation of the carrier frequency is particularly important feature of the SE in the context of this work. We believe that we have found non-trivial manifestation of the solar 5-min oscillations on the Earth.

___________________________________

A. Guglielmi   e-mail: guglielmi@mail.ru
A. Potapov     e-mail: alpot47@mail.ru
B. Dovbnya     e-mail: dovbnya@inbox.ru

## 1. Introduction

It is known that the solar photosphere permanently oscillates with a period of about five minutes (Leighton et al., 1962; Ulrich, 1970). The oscillations cover the entire surface of the Sun. They do not terminate even during the periods when the Sun is in a quiet state. These are the acoustic oscillations, and they are related to the processes of granulation and supergranulation, i.e. formation of convection cells in the solar interior. The review (Vorontsov, Zharkov, 1981) and papers (De Moortel et al., 2002; Aschwanden, 2008; Su et al., 2013) contain a more complete information concerning the permanent 5-minutes solar oscillations.

Now, we want to pay attention to the permanent oscillations in the other frequency band and elsewhere. We are talking about the ion cyclotron electromagnetic waves which are self-excited in the interplanetary medium as a result of plasma instability. At AU distance from the Sun the waves exist in the range of periods 0.2–10 s. According to the standard classification of ultra-low-frequency (ULF) waves of the natural origin (Troitskaya, Guglielmi, 1967; Guglielmi, Pokhotelov, 1996) this range is designated as Pc1–2. (For comparison, the 5-minute solar oscillations fall within the Pc5 range). The existence of ion cyclotron waves in the solar wind follows from the profound and multilateral theoretical studies (Guglielmi, 1979; Melrose, 1986; Gomberoff, 1996; Li et al., 1999; Isenberg et al., 2000; Tu and Marsch, 2001; Kasper et al., 2013) as well as from the detailed analysis of the measurements in situ (Kellogg et al., 2006; Jian et al., 2009, 2010, 2014).

In the present paper our focus will be on the impact of the photospheric 5-minute oscillations on the ion cyclotron waves in the solar wind. At statement of this task, we take into account the fact that the energy of the photospheric oscillations is transformed partly into the energy of Alfven waves that propagate upward and reach the solar corona. Then they are transported into the heliosphere over long distances by the solar wind (Erdélyi et al. 2007; Tomczyk et al. 2007; Mathioudakis et al. 2013). According to Potapov et al. (2012, 2013) the Alfven waves driven by the 5-minute solar oscillations are observed at the Earth's orbit.

We proceed from the assumption that the ion cyclotron waves in solar wind are experiencing modulation with a characteristic period of 5 minutes under the influence of Alfven waves driven by photospheric motions. We shall not try here to present an exhaustive definition of modulation, but will simply imagine a modulated wave as "nearly" monochromatic with parameters (amplitude, phase, frequency) varying in space and in time smoothly and slowly. In our case, this condition certainly holds because the lengths and periods of Alfven waves are much greater than the lengths and periods of ion cyclotron waves.

The paper contains two major sections. In the next section, we present theoretical arguments which testify a deep modulation of the carrier frequency of ion cyclotron waves in the solar wind. After that, we present the search result of quasi-monochromatic ULF electromagnetic oscillations in the range of Pc1–2, the carrier frequency of which varies with the quasi-period of 5 minutes.

**2. Frequency modulation of the ion cyclotron waves**

The theory of solar wind (Parker, 1963) predicts that the pressure of interplanetary plasma is anisotropic, with $p_\perp < p_\parallel$ at the orbit of the Earth, and this prediction is completely confirmed by *in situ* observations (Hundhausen, 1972). Here $p_\parallel$ ($p_\perp$) is the pressure along (across) the interplanetary magnetic field (IMF) lines. At $p_\perp < p_\parallel$ the fire-hose instability may appear, leading to an increase of long-wave hydromagnetic perturbations ($\omega' \ll \Omega_p$), and ion cyclotron short-wave instability ($\omega' \sim \Omega_p$). Here $\omega'$ is the frequency in the co-moving reference system, $\Omega_p = eB/m_p c$ is the proton gyrofrequency, $e$ and $m_p$ are the charge and mass of the proton, $c$ is the velocity of light, $B$ is the modulus of IMF. We know that the threshold of the cyclotron instability is lower, and grows rate is higher than that for the fire-hose instability (Guglielmi, 1979). And we know that adiabatic expansion of the solar corona is in itself a permanent mechanism for the formation of pressure anisotropy of the interplanetary plasma. With increasing the distance $r$ from the Sun the pressure ratio $p_\parallel / p_\perp$ increases rather quickly (approximately as $r^2$). It is supposed that the excitation of ion cyclotron emission due to development of instability leads to the quasi-linear restriction for the pressure anisotropy. This determines the importance of studying the ion cyclotron waves in the interplanetary plasma. But in this paper we would like to explore another aspect of the problem, as it described in general terms in the Introduction.

In the high-beta interplanetary anisotropic plasma with $p_\perp < p_\parallel$ the increment of ion cyclotron instability is estimated as

$$\gamma(k_\parallel, \theta) = \mu(k_* - k_\parallel)\exp\left[-(k_0/k_\parallel)^2\right] - \eta\theta^2, \qquad (1)$$

where $k_* \sim k_0 \sim \omega_{0p}/c$, $\omega_{0p} = \sqrt{4\pi e^2 N/m_p}$, $N$ is the electron concentration, and $\theta$ is the angle between the vectors **B** and **k**, $\theta^2 \ll 1$ (Guglielmi, 1979). We will not give here the formulas for the relation between the values $\mu$, $\eta$ from one hand and the parameters of interplanetary plasma from the other, since they are rather cumbersome. For us it is important only that $\mu > 0$, $\eta > 0$ in a typical case. Therefore, the increment $\gamma(k_\parallel, \theta)$ has a maximum at $\theta = 0$ and $k_\parallel = k_m \sim \omega_{0p}/c$.

Let us use this theoretical prediction for the formulation of an interesting experimental problem. We would like to draw attention to the special property of the cyclotron oscillations when they are observed by using the spacecraft or the ground based magnetometers. This property consists in that the oscillations are experiencing a deep frequency modulation. Indeed, in the laboratory reference system the frequency equals

$$\omega = \mathbf{kU} + \omega'. \qquad (2)$$

Since $k_\perp = 0$ we have $\mathbf{kU} = k_m U \cos\psi$, where **U** is the solar wind velocity, and $\psi$ is the angle between the vectors **U** and **B**. In view of $k_m \approx \omega_{0p}/c$, we obtain the following estimate for the carrier frequency of oscillations

$$\omega \approx \omega_{0p}(U/c)\cos\psi. \qquad (3)$$

Here we ignore the second term on the right hand side of equation (2). This is acceptable in typical cases because $U \gg c\Omega_p/\omega_{0p}$ with a large margin. Exceptions arise from time to time, if $\psi$ is close to the right angle: $|\psi - \pi/2| < (c/U)(\Omega_p/\omega_{0p})$. For example, $|\psi - \pi/2| < 0.1$ when $U \sim 4\cdot 10^7$ cm s$^{-1}$, $\Omega_p \sim 0.5$ s$^{-1}$, and $\omega_{0p} \sim 4\cdot 10^3$ s$^{-1}$. If $\psi = \pi/2$, i.e. the vectors

**U** and **B** are mutually orthogonal, we have $\omega_{min} \sim \Omega_p \sim 0.5$ s$^{-1}$. The maximum frequency

$$\omega_{max} \approx (U/c)\omega_{0p} \sim 5 \text{ s}^{-1} \qquad (4)$$

is reached at $\psi = 0$, when two vectors **U** and **B** are parallel to each other.

So, the theory predicts a deep frequency modulation of ion cyclotron emission. The frequency modulation is expected mainly from variations in the IMF orientation. Besides, the emission should be quasi-continuous, since the pressure anisotropy is a permanent property of the solar wind. Quasi-monochromatic oscillations of the natural origin with a deep modulation of the carrier frequency would be interesting to observe experimentally.

## 3. Serpentine emission

To test the theoretical predictions, first of all we should find the permanent quasi-monochromatic oscillations of natural origin in the Pc1–2 frequency band, the "instantaneous" frequency of which varies with time in a wide range. Of course, the observations onboard a spacecraft would suit best for the search. But we should not neglect also the data of ground-based observations, especially in view of that the electromagnetic oscillations possessing the features indicated above have long been found in the polar caps. Guglielmi and Dovbnya (1973) have discovered such oscillations, and offered to call them "Serpentine Emission" or SE for brevity. Figure 1 shows an example of SE registered at the Vostok station near the South Geomagnetic Pole. We see that the dynamic spectrum has the form of a bending dark band; it really looks like a crawling serpent.

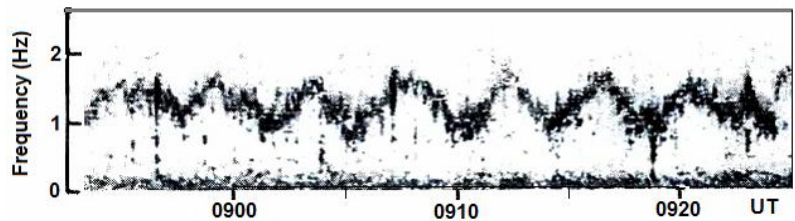

**Figure 1.** The dynamic spectrum of serpentine emission observed on 30 January, 1968 at the Vostok station, Antarctica.

Morphology of the SE is described in the papers (Guglielmi, Dovbnya, 1973, 1974; Fraser-Smith, 1982; Asheim, 1983; Morris, Cole, 1987). We present here the basic information. Electromagnetic oscillations are observed at the high latitude observatories in the frequency range of 0.1–2 Hz. The most important distinguishing feature of SE is a deep frequency modulation. The quasi-period of modulation varies from case to case from ~ 1 min to ~1 h. The amplitude of oscillations lies typically in the interval 0.03–0.3 nT. The oscillations are observed under the quiet geomagnetic conditions and last continuously from hours to days.

The permanency, range of frequencies, and the deep frequency modulation of SE correspond to the qualitative properties of ion cyclotron waves in the solar wind as they have been described in the previous section. Hence, it is reasonable to hypothesize that SE is the ion cyclotron waves of interplanetary origin penetrating into magnetosphere and arriving the Earth at high latitudes. The frequency modulation is mainly due to the variations in orientation of IMF. These variations are created by the large-amplitude Alfvén waves propagating in the interplanetary medium predominantly outward from the Sun within a broad range of wavelengths (Belcher, Davis, 1971).

According to Belcher and Davis (1971) the most Alfvén waves in the interplanetary medium are generated at or near the Sun. Among these waves there are certainly also the quasi-monochromatic 5-minute Alfvén waves which are driven by the photospheric oscillations. Let us assume for a moment that our ideas about the origin of SE are fully correct. Then we should at least to give examples of SE which show clearly the 5-minute modulation of the carrier frequency. To this end we analyzed the available for us archive data of the magnetometric registration of Pc1–2 waves at the high-latitude Antarctic station Vostok. The archive is stored in the Borok Observatory, and it contains the valued data on the ULF waves, which were collected during the Antarctic expeditions organized by Professor Valeria Troitskaya in the sixties of last century. As a result we found the numerous excellent examples testifying in support of our assumption.

We would like to consider Figure 1 once more. We see the serpentine emission of rare beauty. The stability of oscillatory regime is due to the calm conditions in the solar wind and the magnetosphere. The velocity of wind is less than 400 km/s, the interplanetary plasma density is relatively low, the index of geomagnetic disturbance is very moderate, Kp = $2_+$. But the most important fact is that the quasi-period of modulation is approximately 4.5 minutes. It is quite close to the expected value.

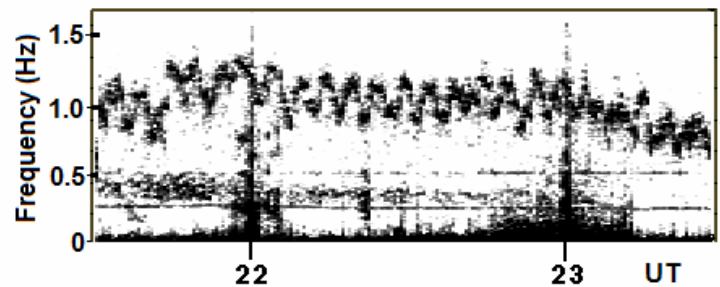

**Figure 2.** The dynamic spectrum of SE with 5-minute modulation of the carrier frequency (Vostok, 15 July, 1968).

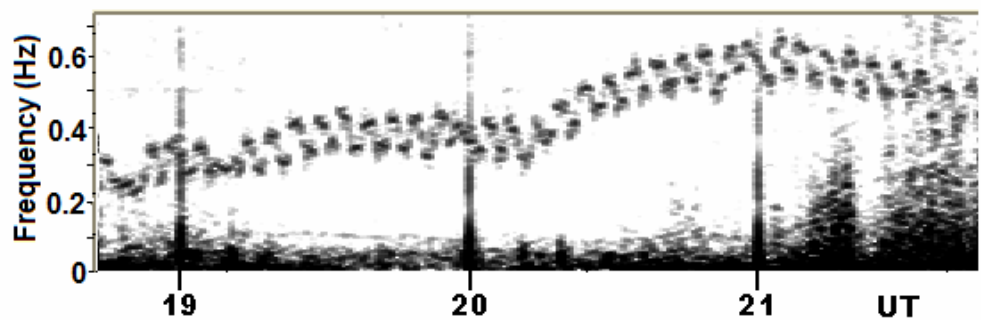

**Figure 3.** The dynamic spectrum of SE with 5-minute modulation on the smooth background of mean frequency changes over a broad range (Vostok, 25 July, 1970).

Clear 5-minute modulation of the SE carrier frequency we see in Figure 2. The averaged frequency is more or less stable in this case. In contrast, Figure 3 shows the case of an almost two-fold change in the frequency for 2 hours. These smooth background changes in mean frequency are evidently related to the nonsinusoidal large-scale Alfvén waves in the solar wind described by Belcher and Davis (1971). It is important that the period of quasi-sinusoidal modulation of the carrier frequency remains equal to 5 minutes in spite of a significant deviation of the mean frequency. This is consistent with our view on the origin of SE.

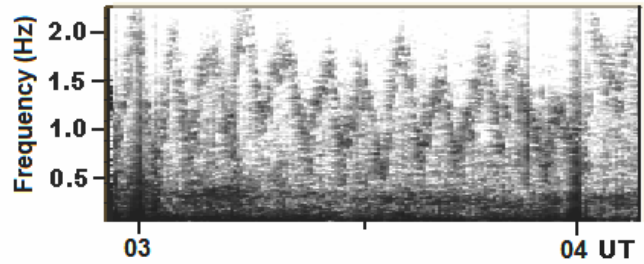

**Figure 4.** The dynamic spectrum of SE showing unusually deep modulation of the carrier frequency (Vostok, 2 August, 1966).

Figure 4 is interesting because it shows the frequency modulated oscillations with a very deep modulation. Swing of the frequency in this example is much greater than the frequency spread we see in Figures 2 and 3. The carrier frequency changes four times, from 0.5 to 2.0 Hz with a period of 5 minutes. If our understanding of the SE nature is correct, then variability in the depth of five-minute modulation indicates instability of conditions of Alfven waves propagation on their way from the solar photosphere to the Earth.

Finally, we give an example of a very long serpentine emission with irregular complex modulation of frequency and variable amplitude and mean frequency (Figure 5). Modulation period is also unstable; in addition to 5 minutes longer modulation periods present. At the same time the magnetosphere is calm, Kp index is not higher than $1_+$. Apparently, there were chaotic variations in $\psi$ angle between the IMF and the solar wind velocity vector.

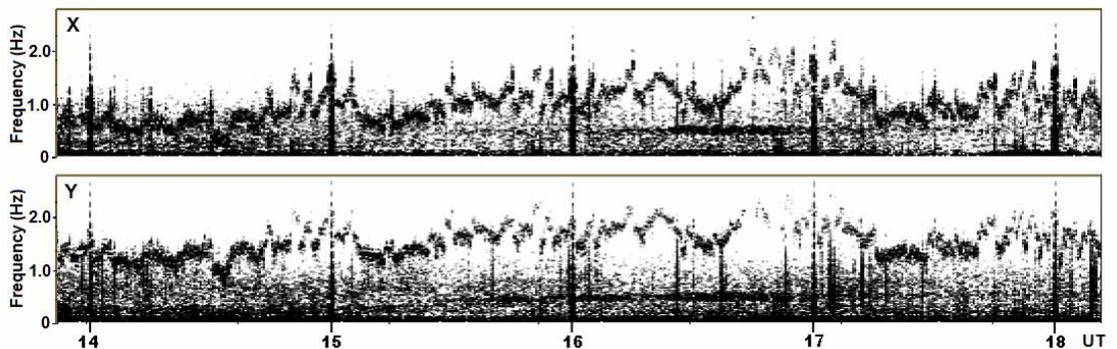

**Figure 5**. The four-hour spectrogram of serpentine emission with irregular behavior of the carrier frequency. Two horizontal components of the emission are shown, *X* and *Y* (Vostok, 19 January, 1966).

So, we found convincing examples of 5-minute modulation of SE carrier frequency. But this is certainly not enough to finally accept the hypothesis of photospheric origin of this modulation. In this regard, we are planning a more comprehensive analysis of all available SE records collected over several years of observations at Vostok station.

## 4. Discussion and Conclusion

We are sure that the theoretical conclusion concerning the frequency modulation of ion cyclotron waves in the solar wind is apodictic. This conclusion is based on fairly simple and physically clear arguments. It could have been objected that the equation (1), which we have used in our arguments, is of limited applicability. Indeed, this equation was derived by linearization of the basic equations of plasma electrodynamics on condition that the plasma and external magnetic field are distributed in space uniformly. Thus, the increment $\gamma$ is indicator of the exponential amplitude growth of infinitesimal weak waves propagating in a homogeneous medium. But exponential growth of amplitude rapidly withdraws our system from the region of applicability of linear theory. Nonlinear processes are developed for a time of about $\gamma^{-1}$, and this restricts the growth of amplitude. In addition the real plasma contains irregularities of various sizes. Scattering by small-scale fluctuations leads to a broadening of the angular spectrum of the waves. The quantitative calculation of nonlinearity and scattering of the waves is a rather complicated task. Here we restrict ourselves to discussion of the qualitative aspects of issue.

For simplicity we assume that the scattering occurs at very small angles. It is natural to believe that the average value of the scattering angle is zero. Let's denote the rms value of the scattering angle as $<\theta^2>$. We will make the averaging of $\gamma(k_\parallel, \theta)$ over $\theta$ and denote the result of averaging as $\gamma(k_\parallel)$. The value $\gamma(k_\parallel)$ is determined by the right-hand side of equation (1) in which $\theta^2$ is replaced by $<\theta^2>$. The term $\eta<\theta^2>$ describes the attenuation of waves due to the scattering. Now we expand the function $\gamma(k_\parallel)$ in the vicinity of its maximum

$$\gamma(k_\parallel) = \gamma_m \left[1 - (k_\parallel - k_m)^2 / \Delta k^2 \right] \tag{5}$$

Here $\gamma_m$ is the maximum of the increment, and $\Delta k$ is the width of the instability range. The values $\gamma_m$, $k_m$ and $\Delta k$ are determined by the parameters $\mu$, $\eta$, $k_*$, $k_0$, and $<\theta^2>$, but we will not dwell on this, since we are only interested in the overall structure of equation (5).

With taking into account equations (2), (5) the evolution of small perturbations looks as follows in the laboratory reference system. The spectral components will be amplified if $|k_\parallel - k_m| < \Delta k$. The component with $k_\parallel = k_m$ tends to dominate with time, and it would be possible to observe the monochromatic ion cyclotron wave with frequency $\omega_m = k_m U \cos\psi$. The wave amplitude increases exponentially with growth rate $\gamma_m$ according to the linear theory. But with time the linear approximation ceases to work due to nonlinear modification of the medium under the action of ponderomotive forces (Lundin and Guglielmi, 2006). The nonlinear modification leads to the modulational instability of the monochromatic wave. In its turn, the self-modulation of the ion cyclotron waves leads to a broadening of the wave spectrum. This opens the channel for energy transfer from the band of wave amplification $|k_\parallel - k_m| < \Delta k$ into the band of wave dissipation $|k_\parallel - k_m| > \Delta k$. The larger the wave amplitude, the sink of wave energy is more intensive. As a result, the saturation of the wave amplitude occurs. It should be noted that this fairly general mechanism of the amplitude stabilization of the waves in unstable medium was borrowed from the theory of wind waves on the water (Stepanyants and Fabrikant, 1989).

So, the theoretical conclusion regarding the deep frequency modulation of ion cyclotron waves under the influence of Alfven waves driven by photospheric motions is reliable enough, because it rests on the analysis of simple and rough properties of the oscillatory system. The situation is different with the question of experimental observation of the frequency-modulated quasi-monochromatic waves in the range of Pc1–2. It would seem that the properties of SE which described in the previous section of this paper are consistent with the theoretical predictions. But we have poor understanding of how the ion cyclotron waves penetrate from the interplanetary medium into the Earth's magnetosphere. Besides, the morphology of SE is studied insufficiently in order to reliably draw conclusions concerning origin of this emission. In addition to the first publications by Guglielmi and Dovbnya (1973, 1974) just a few papers by other authors are known to be devoted to the observations of SE (Troitskaya, 1979; Fraser-Smith, 1982; Asheim, 1983; Morris and Cole, 1986, 1987). Let us consider shortly these papers.

The SE has been detected in the both polar caps. Troitskaya (1979) and Fraser-Smith (1982) have presented the reviews of observations at the Vostok station in Antarctic. Morris and Cole (1986, 1987) have studied the morphology of SE according to the observations at the Antarctic station Davis. They investigated all the main characteristics of the SE. Qualitative agreement of the results of observation at Vostok and Davis is of no doubt. The observations of SE in Arctic were made by Asheim (1983) at the station Ny-Ålesund, Spitsbergen. We should also mention the episodic registration of SE at the drifting ice station North Pole-22. An example of such registration is presented in the monograph (Guglielmi, 1979). The authors of all these works agree on the fact that SE is an interesting natural phenomenon, and they do not exclude extra-magnetospheric origin of this emission. But none of the previous papers made any mentions of relation between the modulation of SE and oscillations of the solar photosphere.

Morris and Cole (1987) have pointed to a number of reasons why SE does not attract a wide attention of researchers. Perhaps their list should be supplemented by another consideration. We have in mind the doubts related to uncertainty of the origin of such unusual oscillations. But we still hope that our paper will stimulate further researches that will shed light on the origin of SE, and will allow us to better understand the role of SE in the overall system of solar-terrestrial relations.

**Funding:** The work was supported by the Russian Foundation for Basic Research (projects 13-05-00066 and 13-05-00529).
**Conflict of Interest:** The authors declare that they have no conflict of interest.

**Figure legends**

Figure 1. The dynamic spectrum of serpentine emission observed on 30 January, 1968 at the Vostok station, Antarctica.

Figure 2. The dynamic spectrum of SE with 5-minute modulation of the carrier frequency (Vostok, 15 July, 1968).

Figure 3. The dynamic spectrum of SE with 5-minute modulation on the smooth background of mean frequency changes over a broad range (Vostok, 25 July, 1970).

Figure 4. The dynamic spectrum of SE showing unusually deep modulation of the carrier frequency (Vostok, 2 August, 1966).

Figure 5. The four-hour spectrogram of serpentine emission with irregular behavior of the carrier frequency. Two horizontal components of the emission are shown, *X* and *Y* (Vostok, 19 January, 1966).